\documentclass[twocolumn,letterpaper,amsmath,amssymb,floatfix,aps,superscriptaddress]{revtex4} %,showpacs,

\usepackage{graphicx}% Include figure files
\usepackage{dcolumn}% Align table columns on decimal point
\usepackage{bm}% bold math
\usepackage{epsfig,color}

\newcommand{\Av}[1]{{\mathbf #1}}

\newcommand {\bnabla} {\mbox{\boldmath$\nabla$}}
\newcommand {\bomega} {\mbox{\boldmath$\omega$}}

\def\ul#1#2{\textstyle{\frac#1#2}}

\def\ln{{\operatorname{ln}}}

\def\rmd{{\mathrm{d}}}
\def\rmi{{\mathrm{i}}}
\def\rme{{\mathrm{e}}}

\begin{document}

%\title{Generalizations and ramifications induced by the counterion structure in the strong and weak-coupling electrostatic interactions}
% Role of Quadrapoles in Interactions between Charged Surfaces Through Water:  Strong and Weak Coupling
\title{Role of Multipoles in Counterion-Mediated Interactions between Charged Surfaces:  Strong and Weak Coupling}
%\title{Role of Multipoles in Counterion-Mediated Interactions between Charged Surfaces}

\author{M. Kandu\v c}
\affiliation{Department of Theoretical Physics,
J. Stefan Institute, SI-1000 Ljubljana, Slovenia}

\author{A. Naji}
\affiliation{Department of Physics, Department of Chemistry and Biochemistry, \&
Materials Research Laboratory,
University of California, Santa Barbara, CA 93106, USA}
\affiliation{Kavli Institute of Theoretical Physics, University of California, Santa Barbara, CA 93106, USA}
\affiliation{School of Physics, Institute for Research in Fundamental Sciences (IPM), P.O. Box 19395-5531, Tehran, Iran}

\author{Y.S. Jho}
\affiliation{Materials Research Laboratory, University of
California, Santa Barbara, CA 93106, USA}
\affiliation{Dept. of Physics, Korea Advanced Institute of Science and Technology,
Yuseong-Gu, Daejeon, Korea 305-701}

\author{P.A. Pincus}
\affiliation{Materials Research Laboratory, University of
California, Santa Barbara, CA 93106, USA}
\affiliation{Dept. of Physics, Korea Advanced Institute of Science and Technology,
Yuseong-Gu, Daejeon, Korea 305-701}

\author{R. Podgornik}
\affiliation{Department of Theoretical Physics,
J. Stefan Institute, SI-1000 Ljubljana, Slovenia}
\affiliation{Kavli Institute of Theoretical Physics, University of California, Santa Barbara, CA 93106, USA}
\affiliation{Institute of Biophysics, Medical Faculty and Department of Physics, Faculty of Mathematics and Physics,
University of Ljubljana, SI-1000 Ljubljana, Slovenia}

\begin{abstract}
We present general arguments for the importance, or lack thereof, of the structure in the charge distribution of counterions for counterion-mediated interactions between bounding symmetrically charged surfaces. We show that on the mean field or weak coupling level, the charge quadrupole contributes the lowest order modification to the contact value theorem and thus to the intersurface electrostatic interactions. The image effects are non-existent on the mean-field level even with multipoles. On the strong coupling level the quadrupoles and higher order multipoles contribute additional terms to the interaction free energy only in the presence of dielectric inhomogeneities. Without them, the monopole is the only multipole that contributes to the strong coupling electrostatics. We explore the consequences of these statements in all their generality.
\end{abstract}
\maketitle

\section{Introduction}

The assembly of colloidal building blocks with designer engineered size,  shape, and chemical anisotropy seems to be the next step in the fundamental and applied colloid science and science of soft materials, vigorously pursued by many researchers \cite{vanBlaaderen,hong}.  Moving beyond the traditional systems implies the creation of designer colloidal building blocks such as colloidal molecules, janus spheres, and other patchy particle motifs that have very different propensities for self-assembly \cite{Glotzer}. In order to realize this goal one needs detailed control over various aspects of colloid geometry at the nanoscale ({\em e.g.}, aspect ratio, faceting, branching, roughness) and microscale ({\em e.g.}, chemical ordering, shape gradients, unary and binary colloidal ``molecules'').  The ability to spatially modify the surface structure of colloids  with designed chemical heterogeneity, {\em e.g.}, to form a patchy surface structure, seems to be becoming a realistic goal. Fabrication of stable anisotropic microcapsules was recently accomplished by the layer-by-layer polyelectrolyte adsorption technique combined with particle lithography technique to produce anisotropic polymer microcapsules with a single nanoscale patch \cite{huda}. Granick and co-workers \cite{hong} recently reported a highly scalable synthetic pathway for creating bipolar janus spheres, {\em i.e.}, particles that consist of oppositely charged hemispheres.  Such colloidal ``animals" which are probably the simplest example of patchy colloids, exhibit orientation-dependent interactions that go together with localized patches of like/unlike charges or hydrophobic/hydrophilic regions. This heterogeneous interaction landscape promotes the formation of larger colloidal molecules and clusters that are themselves patchy.

These advances in the nano- and microscopic tailoring of (charged) colloids motivated various approaches to generalizations of the existing theories of electrostatic interactions in charged colloids by explicitly including the structure of the counterions as embodied by their multipolar moments. The inclusion of structured counterions of a dipolar \cite{abrashkin,bohinc1} and quadrupolar  \cite{bohinc,woon}  type into the theory of electrostatic colloidal interactions has brought fourth some of the salient features of the counterion structure effects, which on face value appear to be quite distinct from the standard
Poisson-Boltzmann framework. In view of these advances in the study of
interactions between charged colloids it thus seems appropriate to explore the ramifications of the emerging paradigm if applied to these more complicated structured colloidal molecules. Especially the non-spherically symmetric charge distribution of microscopically tailored colloidal particles might lead to some unexpected properties of electrostatic interactions in this type of systems. With this in mind we thus embark on a thorough examination of the consequences of multipolar charge distribution of mobile counterions that mediate interactions between charged (planar) macroions.

\section{Model}

\begin{figure}[t]
%\centerline{\psfig{figure=geometry.eps,width=7cm}}
\centerline{\psfig{figure=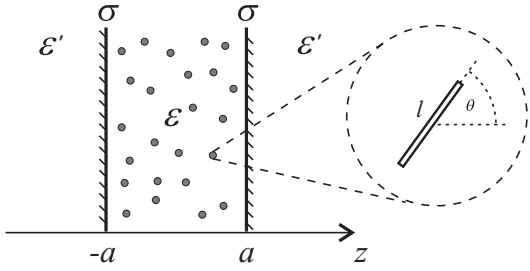,width=7cm}}
\caption{Geometry of the system. Two charged walls with different dielectric constants and structured counterions in between. Counterions are treated as pointlike particles but with internal uniaxial structure that can be viewed, for example, as uniformly charged rods.}
\label{geom}
\end{figure}

In what follows we will consider a system of fixed macroions of surface charge distribution $\rho_0(\Av r)$ in an aqueous solution, described as a dielectric continuum with a dielectric constant $\varepsilon$ and temperature  $T$, containing
$N$ neutralizing counterions. Counterions are assumed to be pointlike particles but they do posses a rigid internal structure described by a charge distribution $\hat \rho(\Av r; {\Av R}_i, \bomega_i)$ that we assume can be written as a standard multipolar expansion \cite{schwinger}
\begin{eqnarray}
\hat \rho(\Av r; {\Av R}_i, \bomega_i) &=& e_0q \delta(\Av r - {{\Av R}_i}) - p_0 (\Av n_i \cdot \bnabla)\delta(\Av r - {{\Av R}_i})+ \nonumber\\
&&+ t_0 (\Av n_i \cdot \bnabla)^2 \delta(\Av r - {{\Av R}_i}) + \cdots,
\label{chaden}
\end{eqnarray}
for the $i$-th counterion located at ${\Av R}_i$ with $\bomega_i$ being the orientational variables specifying the angular dependence of the counterion charge density. The monopolar moment of each counterion is $e_0 q$, where $q$ is the charge valency and $e_0$ the elementary charge, $\Av p = p_0~ \Av n$ is the dipolar moment and $\Av Q = t_0 ~ \Av n \otimes \Av n$ the quadrupolar moment with director unit vector $\Av n$. We must emphasize here that the quadrupolar expansion in (\ref{chaden}) is not general but adequately describes only a uniaxial counterion, {\em e.g.}, a charged particle of rod-like structure, Fig.~\ref{geom}.
One possible implementation of counterions possesing just monopolar and quadrupolar moment is a uniformly charged rod with charge $e_0 q$ and length $l$. Counterion's quadrupolar moment is then $t_0= e_0 q l^2/24$.
Another possibility is a negative charge ($-e_2$) in the center and two positive charges ($+e_1$) located at both ends of the rod leading to monopolar and 
quadrupolar moments of $e_0 q=2e_1-e_2$ and $t_0=e_1 l^2$, respectively.

Counterions interact {\em via} a Coulomb interaction potential $u(\Av r,\Av r')$ so that the interaction energy of two given counterions $i$ and $j$ at set positions and orientational configurations is obtained by integrating the Coulomb interaction over  their internal orientational degrees of freedom as
\begin{equation}
\int\!\!\!\! \int \rmd\Av r' \rmd\Av r \hat \rho(\Av r; {\Av R}_i, \bomega_i)u(\Av r,\Av r') \hat \rho(\Av r'; {\Av R}_j, \bomega_j).
\end{equation}
Although our formalism in this section is in general applicable to macroions of arbitrary shape and charge distribution, we shall primarily focus on the case of two charged planar surfaces located at $z=\pm a$ with the charge distribution
\begin{equation}
\rho_0(\Av r)=\sigma\delta(a-z)+\sigma\delta(a+z).
\label{sigmas}
\end{equation}
We also consider a dielectric inhomogeneity between the bounding surfaces and the ionic solution, such that the ionic solution is described with dielectric constant $\varepsilon$ and the bounding surfaces with $\varepsilon'$. In this geometry the Coulomb interaction potential is composed of the direct interaction $u_0(\Av r,\Av r')=(4\pi \varepsilon\varepsilon_0 |\Av r - \Av r'|)^{-1}$ and the (electrostatic) image interaction $u_{\rm im}(\Av r,\Av r')$, so that
\begin{equation}
u(\Av r,\Av r')=u_0(\Av r,\Av r')+u_{\rm im}(\Av r,\Av r').
\label{decomp-u}
\end{equation}
The Green's function in planar geometry can be expressed as a sum of image charge contributions, {\em viz.}
\begin{eqnarray}
%\hspace{-3ex}
u(\Av r,\Av r')=\frac{1}{4\pi\varepsilon\varepsilon_0}&&\left[\sum_{n\textrm{ even}}
\frac{\Delta^{|n|}}{|\Av r'-\Av r-2na\, {\hat{\Av k}}|}\nonumber\right.+\\
&&\left.+\sum_{n\textrm{ odd}}\frac{\Delta^{|n|}}{|\Av r'-\Av r+(2na-2z')\, {\hat{\Av k}}|}\right].
\label{green}
\end{eqnarray}
In both sums the index $n$ runs through negative and positive integer values. The term corresponding
to $n=0$ represents the direct Green's function of the particle in a space of a uniform dielectric constant, namely $u_0(\Av r,\Av r')$.
The unit vector $\hat{\Av k}=(0,0,1)$ points in the $+z$ direction. The relative dielectric jump at the two bounding surfaces is quantified as $\Delta=(\varepsilon-\varepsilon')/(\varepsilon+\varepsilon')$. For most relevant situations the outer permittivity is smaller than the inner one, $\varepsilon'<\varepsilon$, so that the dielectric jump is positive, $\Delta>0$ and vanishes in the homogeneous case with $\varepsilon' \rightarrow \varepsilon$.

The canonical partition function $Z_N$ of this Coulomb fluid composed of $N$ charged particles is given by
\begin{eqnarray}
Z_N&=&\frac{1}{N!}\int  \rmd\Av R_1\cdots \rmd\Av R_N \rmd\Av \bomega_1\cdots \rmd\Av \bomega_N\>\times\\
&&\hspace{4ex}\times\exp\Bigl(-\beta U[{\Av R}_i, \bomega_i; {\Av R}_j, \bomega_j] \Bigr),
\end{eqnarray}
with the configurationally dependent ionic interaction energy given by
\begin{eqnarray}
&&\hspace{-4ex}U[{\Av R}_i,\bomega_i; {\Av R}_j, \bomega_j] =\nonumber\\
&&\frac{1}{2}\sum_{i\ne j}\int\!\!\!\!\int \rmd\Av r \rmd\Av r'
\hat \rho(\Av r; {\Av R}_i, \bomega_i)\,u(\Av r,\Av r')\, \hat \rho(\Av r'; {\Av R}_j, \bomega_j) + \nonumber\\
& &
+ \sum_i \int\!\!\!\!\int \rmd\Av r \rmd\Av r' \hat \rho(\Av r; {\Av R}_i, \bomega_i)\,u(\Av r,\Av r')\rho_0(\Av r') + \nonumber\\
& & + \frac{1}{2} \int\!\!\!\!\int \rmd\Av r \rmd\Av r' \rho_0(\Av r)\, u(\Av r,\Av r')\,\rho_0(\Av r'),
\end{eqnarray}
where $\beta=1/k_{\mathrm{B}} T$.  The three terms in $U[{\Av R}_i,\bomega_i; {\Av R}_j, \bomega_j] $ correspond to direct electrostatic interaction between counterions, electrostatic interactions between counterions and fixed charges and between fixed charges on the walls themselves, respectively. We will furthermore make the standard assumption that the system is overall electroneutral, implying that the mobile countercharge exactly compensates the charge on the surfaces.

We now proceed in the Netz \cite{Netz} fashion and perform the Hubbard-Stratonovitch transformation of the partition function \cite{podgor}, where the configurational integral over counterion positions is transformed into a functional integral over a fluctuating auxiliary electrostatic potential $\phi(\Av r)$. In this way the grand-canonical partition function is obtained straightforwardly in the form \cite{Netz}
\begin{equation}
Z =\sum_{N=0}^\infty \lambda^N Z_N = {\cal C}\int{\cal D}[\phi(\Av r)]\,\rme^{-\beta H[\phi(\Av r)]},
\label{funint}
\end{equation}
where $\lambda$ is the bare fugacity which is the exponential of the chemical potential. The prefactor $\cal C$ above is the functional determinant of the inverse Coulomb kernel $u^{-1}(\Av r,\Av r')$ while the ``action'' of the functional integral is given by
\begin{eqnarray}
H[\phi]&=&\!\!\int\!\!\rmd\Av r \bigg[{\textstyle\frac{1}{2}}\varepsilon(\Av r)\varepsilon_0
(\bnabla\phi)^{2} -\rmi\rho_0(\Av r)\phi(\Av r) - \nonumber\\
&&\hspace{-4ex}-\frac{\lambda'}{\beta}\!\int\!  \rmd \bomega ~\Omega(\Av r)\exp{\left(\! \rmi\beta \!\!\int \!\! \rmd\Av r'\hat \rho(\Av r'; \Av r, \bomega)\phi(\Av r')\!\right)} \bigg].
\label{action}
\end{eqnarray}
Here we have introduced the renormalized fugacity as $\lambda'=\lambda\,\exp(\frac{1}{2}  u_0(\Av r,\Av r))$, where $u_0(\Av r,\Av r)$ is the electrostatic direct self-energy of a single counterions with multipolar charge distribution. $\Omega(\Av r)$ is the geometric characteristic function of the counterions, being equal to unity in the slab between the bounding surfaces and zero otherwise.
The partition function in the above form can not be evaluated explicitly except in the one-dimensional case \cite{dean} where the partition function evaluation is reduced to a solution of a Schr\" odinger-like equation. Nevertheless the field-theoretical representation of the grand canonical partition function allows one to use quite powerful analytical approaches that eventually lead to an explicit evaluation of the partition function in two well defined and complementary limits \cite{Netz,Naji,attard} that we shall address later in this paper. These limits retain the relevance also in the case of structured counterions.

%%%%%%%%%%%%%%%%%%%%%%%%%%%%%%%%%%%%%%%%%%%%%%%%%%%%%%%%%%%%%%%%%%%%%%%%%%%%%%%%%%%%%%%%%%%%%%%%
\section{Dimensionless representation}
\label{sec:rescal}

One may obtain a dimensionless representation for the present system by rescaling all length scales with 
a given characteristic length scale. Recall that the characteristic distance at which two unit charges interact with thermal energy $k_{\mathrm{B}}T$
is known as the Bjerrum length $\ell_{\mathrm{B}}=e_0^2/(4\pi\varepsilon\varepsilon_0 k_{\mathrm{B}}T)$ (in water at room temperature, the value is $\ell_{\mathrm{B}}\approx 0.7$ nm).
If the charge valency of counterions is $q$ then the aforementioned distance scales as $q^2 \ell_{\mathrm{B}}$. Similarly, the distance at which a counterion interacts with a macromolecular
surface of surface charge density $\sigma$ with an energy $k_{\mathrm{B}}T$ is called the Gouy-Chapman length, defined as $\mu=e_0/(2\pi q\ell_{\mathrm{B}}\sigma)$.
A competition between ion-ion and ion-surface interactions can thus be quantified by the ratio ($\Xi$) of these characteristic lengths, that is 
$$\Xi=q^2 \ell_{\mathrm{B}}/\mu=2\pi q^3 \ell_{\mathrm{B}}^2\sigma/e_0,$$ which is referred to as the electrostatic coupling parameter.

In what follows, we  may rescale the length scales with the Gouy-Chapman length, {\em i.e.}, $\Av r\rightarrow \Av r/\mu$; hence, the surface separation will be rescaled as
$  \tilde D = D/\mu$ or the rescaled half-distance as $\tilde a = a/\mu$. 
Other dimensionless quantities  can be defined as follows. The dimensionless multipolar moments are defined as
\begin{equation}
p=p_0/e_0q\mu \quad{\textrm{and}}\quad t=t_0/e_0q\mu^2,
\end{equation}
and the dimensionless pressure as
\begin{equation}
%\tilde n(z) = \frac{n(z)}{2\pi\ell_{\mathrm{B}} (\sigma_1/e_0)^2}\quad{\textrm{and}}\quad
	\tilde P = \frac{\beta P}{2\pi\ell_{\mathrm{B}} (\sigma/e_0)^2}.
	\label{P_dimless}
\end{equation}

For simple structureless counterions the functional integral (\ref{funint}) can be rescaled \cite{Netz} yielding $$ Z \rightarrow {\cal C}\int{\cal D}[\phi(\Av r)]\,\rme^{- H'[\phi(\Av r)]/\Xi},$$ with dimensionless $H'[\phi(\Av r)]$. As will be shown later an identical representation can be obtained also for structured counterions, described on the level of a multipolar expansion of their charge density.

\section{Strong and Weak Coupling Dichotomy}

In the absence of a general approach that would cover thoroughly all the regions of the parameter space one has to take recourse to various partial formulations that take into account only this or that facet of the problem  \cite{hoda}. The traditional approach to these one-component Coulomb fluids has been the mean-field Poisson-Boltzmann (PB) formalism applicable at weak surface charges, low counterion valency and high temperature \cite{Oosawa, Ohnishi, Naji}. The limitations of this approach become practically important in highly-charged systems where counterion-mediated interactions between charged bodies start to deviate substantially from the mean-field accepted wisdom \cite{hoda}. One of the most important recent advances in this field has been the systematization of these non-PB effects based on the notions of {\em weak} and {\em strong} coupling approximations. They are based on the field-theoretical representation of grand canonical partition function (\ref{action}), whose behavior depends on a single dimensionless coupling parameter $\Xi$ \cite{Netz} (see below). The weak-coupling (WC) limit of $\Xi\rightarrow 0$ coincides exactly with the mean-field Poisson-Boltzmann theory \cite{podgornik} and is based on a collective description of the counterion density. The strong coupling (SC) limit of $\Xi\rightarrow \infty$ is diametrically opposite and corresponds to a single particle description. It has been pioneered by Rouzina and Bloomfield \cite{Rouzina}, elaborated later by Shklovskii {\em et al.} \cite{shklovskii,Nguyen1,Nguyen2} and Levin {\em et al.} \cite{Levin}, and eventually brought into a final form by Netz {\em et al.} \cite{Netz,hoda,Naji, NetzJoanny1,NetzJoanny2,strong}.

These two limits are distinguished by the pertaining values of the coupling parameter $\Xi$ which can be introduced in the following way.

The regime of ${\Xi}\ll 1$ is the case of low counterionic valency and/or weakly charged surfaces, and is referred to as  the weak-coupling limit. It is characterized by the fact that the width of the counterion layer $\mu$ is much larger than the separation between two neighbouring counterions in solution and thus the counterion layer behaves basically as a three-dimensional gas. Each counterion in this case interacts with many others and the collective mean-field approach of the Poisson-Boltzmann type is completely justified.

On the other hand in the strong coupling regime ${\Xi}\gg 1$, which is true for high valency of counterions and/or highly charged surfaces. In this case the mean distance between counterions, $a_\bot \sim \sqrt{e_0q/\sigma}$, is much larger than the
layer width ({\em i.e.}, $a_\bot/\mu\sim \sqrt{\Xi}\gg 1$), indicating that the counterions are highly localized laterally and form a strongly correlated quasi-two-dimensional layer next to a charged surface. In this case, the weak-coupling approach breaks down
due to strong counterion-surface and counterion-counterion correlations. Since counterions can move almost independently from the others along the direction perpendicular to the surface, the collective many-body effects that enable a mean-field description are absent, necessitating a complementary SC description \cite{Netz, Jho, Jho-preprint}.

These two approximations allow for an explicit and exact treatment of charged systems at two disjoint limiting conditions whereas the parameter space in between can be analysed only approximately and is mostly accessible solely {\em via} computer simulations. As will become clear in what follows the WC and SC limits remain valid also for structured counterions described with a multipolar charge distribution.

%\section{Monte Carlo simulations}
%Monte Carlo simulations are performed to test the theoretical predictions. The electrostatic potential is implemented using image charge method as described in \cite{MC}. Each quadrupole is composed of three charged particles which are arranged on a line, $e_1\delta(x-l)-e_2\delta(x)+e_1\delta(x+l)$. The quadrupolar moment of the model particle is then $t_0=e_1l^2$.

%So, we have freedom to choose the $e_2$. In the strong coupling regime, $e_1$ and $e_2$ are chosen to be $2e_1-e_2 = 1$, {\em i.e.} the monopole moment is $1$. Although the $e_1$ or $e_2$ are high, the net positive monopole enforces a strong correlation between the charged quadrupoles and they are well separated. But in the weak coupling regime, the thermal fluctuations easily overcome the monopolar repulsion and the highly charged positive and negative components of quadrupole are able to collapse by chance. To avoid the unwanted collapses, we need to set $e_2$ to be zero and choose small value of $e_1$. For this reason we only can deal small values of $t_0$.

%The charges are considered as point particles. They interact with each other {\em via} electrostatic interactions. There are hard wall boundaries at the surface of the membrane and the particles are not able to penetrate. Dielectric constants are chosen to be $\varepsilon = 79$ and $\varepsilon' = 2$, corresponding to $\Delta=0.95$. Temperature is $300$~K.

\section{Mean-Field Limit}

The mean-field limit, $\Xi\rightarrow 0$, is defined {\em via} the saddle-point configuration of the Hamiltonian (\ref{action}) \cite{podgornik} and is valid for a weakly charged system. This leads to the following generalization of the Poisson-Boltzmann equation
\begin{eqnarray}
\varepsilon\varepsilon_0 \nabla^2\psi_0 (\Av r)&=&  -\lambda'\!\!\int\!\!\!\!\int \rmd\Av r'~\rmd \bomega ~\hat \rho(\Av r; \Av r', \bomega)\,\Omega(\Av r')\times\nonumber\\
&\times&\exp{\left[- \beta \!\int \rmd\Av r'' \hat \rho(\Av r''; \Av r', \bomega)\psi_0(\Av r'') \right]}
\label{PB}
\end{eqnarray}
for the real-valued potential field $\psi_0 = -\rmi \phi_0$.
Note that the dielectric discontinuity at the boundaries is irrelevant within the mean-field theory \cite{matej1} since in the planar geometry this is effectively one-dimensional theory. Taking now the counterion density function as a sum of the monopolar, dipolar and quadrupolar terms, where the mean-field depends only on the transverse coordinate $z$, the above equation can be written in dimensionless form as
\begin{eqnarray}
\psi'' &=& -\frac12 {\textstyle}\int_{-1}^{+1} \!\!\rmd x ~\Omega(x, z)\left(
u(z) - p x u'(z) + t x^2 u''(z) \right),\nonumber\\
&\propto& \rho_1(z) + \rho_2(z) + \rho_3(z),
\label{quadPB1}
\end{eqnarray}
where we have defined dimensionless potential $\psi=\beta e_0 q\psi_0$ with corresponding derivatives $\psi'=\beta e_0 q\mu\psi_0'$ and  $\psi''=\beta e_0 q\mu^2\psi_0''$ and dimensionless multipolar moments $p$ and $t$. We defined the orientational variable $x = \cos{\theta}$,  where $\theta$ is the angle between the $z$-axis, assume to coincide with the normal to the bounding surfaces, the director of the uni-axial counterion is $\Av n$ and the integral over this variable gives the orientational average.
We have also introduced $$ u(z) = C \exp{\left(- \psi - p x \psi' - tx^2 \psi''\right)},$$ which is obviously the local orientationally dependent number density of the counterions and $\rho(z)$ is the corresponding orientationally averaged charge density of the counterions. In the case of higher multipoles the number density and the charge density of the counterions are not proportional. Expressions $\rho_i(z)$ are simply the orientationally averaged multipolar charge densities, {\em i.e.}, $i=1$ for monopolar charge, $i = 2$ for dipolar charge etc., that are simply proportional to the three terms in the integrand of (\ref{quadPB1}). The corresponding prefactor is simply obtained by appropriate normalisation of $\rho_1$ to satisfy electro-neutrality condition of the system.

The above PB equation has to be supplemented by an appropriate boundary condition at $z=\pm a$ corresponding to the
electroneutrality of the system by taking into account surface charges, $\sigma$. The constant $C$ is set by these boundary conditions, as is the case in the standard PB theory. In dimensionless units these boundary conditions read $\psi'(\pm a)=\mp 2$.
Obviously in the case of counterions with only monopolar charge distribution the above set of equations reduces to the standard Poisson-Boltzmann theory.

The characteristic function $\Omega(x, z)$ for the parallel plane geometry simply excludes the counterion configurations that would penetrate the bounding walls and depends on the geometric form of the counterions. Since in what follows we will not be interested in steric effects, we will assume that all counterions are pointlike, and so $\Omega(x, z) = 1$. Steric effects have been studied elsewhere \cite{bohinc,bohinc1,woon}.

% MK
%An alternative way to write Eq. \ref{quadPB1} in the pointlike counterion limit would be
%\begin{equation}
%\varepsilon\varepsilon_0 \psi_0'' =  -
%q  \mathopen< u(z)\mathclose>_x - p  \mathopen< x u'(z)\mathclose>_x + t  \mathopen< x^2 u''(z) \mathclose>_x
%\label{quadPB2}
%\end{equation}
%where the angular average over $x = \cos{\theta}$ is indicated by $\mathopen< \dots \mathclose>_x$.

Because of the similarity with the standard PB equation for monopolar charges one is led to believe that the pressure in an inhomogeneous system of multipolar countercharge can be derived in the same way as in the standard PB theory, {\em via} the s.c. contact value theorem \cite{Andelman}. Indeed this can be proven exactly, namely the mean-field pressure can be obtained from the first integral of the PB equation (\ref{quadPB1}). After some manipulations its first integral is obtained in the form
\begin{equation}
\tilde P=-\tfrac14 {\psi'}^2  + \tfrac14 \!\int_{-1}^{\,1}\!\!\! \rmd x
\left[ u+  p x u\psi' + t x^2 (u\psi''  - u'\psi')\right].
\label{P1}
\end{equation}
Here $\tilde P$ is the dimensionless equilibrium pressure in the system, defined as (\ref{P_dimless}). % is defined as $\tilde P=P/(\sigma^2/2\varepsilon\varepsilon_0)$.
The r.h.s. of the above identity can be calculated at any arbitrary point $|z_0|<a$ as its value is  independent of $z_0$. The choice of the actual point is governed by the symmetry of the system, as is usual also in the standard PB theory. Since we delve only on the symmetrical solution, with inversion symmetry centered on $z = 0$, the derivative of the mean potential at the mid-point must vanish, {\em i.e.}, $\psi'(z_0=0) = 0$. Thus in this case
\begin{equation}
\tilde P=\tfrac14 \int_{-1}^{\,1}\rmd x\,u(0)\left(1 - tx^2 |\psi''(0)|\right).
\label{P3}
\end{equation}
While the first - van't Hoff - term corresponds to repulsive interactions of the standard PB type, the second term in this particular geometry and symmetry entails attractive interactions between the bounding surfaces. Note that $\psi''<0$.

%
%If one chooses $z_0$ as the midpoint of the slab $|z|<a$, the pressure assumes two very different forms, depending on the symmetry of the solutions. If the system is antisymmetric, then $\psi_0(z=0) = \psi_0''(z=0) = 0$. It thus follows straightforwardly in this case that
%\begin{eqnarray}
%\beta p_0 &=&-\frac{\varepsilon\varepsilon_0}{2} [\psi_0'(z=0)]2  + \nonumber\\
%& & + {\textstyle\frac12}\int_{-1}^{+1}\!\! dx
%\left( u(z=0) + p x~ \vert\psi_0'(z=0)\vert u(z=0) + t x^2 \vert\psi_0'(z=0)\vert u'(z=0))\right).\nonumber\\
%~
%\label{P2}
%\end{eqnarray}
%Obviously all three multipolar terms contribute to the pressure in this case.

%Here only the monopolar and the quadrupolar terms contribute to the pressure and we again denoted the orientational average by $\left< \dots \right>_x$. Alternatively, by taking into account the PB equation, one can write the pressure as
%\begin{equation}
%\beta p_0 = \left< u(z=0)\right>_x -  \frac{t q}{\varepsilon\varepsilon_0} \left<u(z=0)\right>_x\left<x^2u(z=0)\right>_x.
%\label{P4}
%\end{equation}
Let us take a closer look at the above expression. For a polyelectrolyte chain, {\em i.e.}, an extended and flexible counterion, one can derive a similar type of pressure formula \cite{rudibridge}, with a  negative term that contributes an attractive part to the force equilibrium. This attractive part in the case of polyelectrolytes  is due to polyelectrolyte bridging interactions \cite{licer} that stem from the connectivity and flexibility of the polyelectrolyte chain.
%The pressure is independent of constant $C$ once the potential $\psi$ is calculated out of Eq. (\ref{quadPB1}), nevertheless that $C$ explicitely occurs in (\ref{P3}).

Does equation (\ref{P3}) have a similar physical content? This interpretation certainly does not seem likely for point counterions, for which the above pressure formula was derived. The structure of the first integral of the PB equation (\ref{P3}) seems to be saying that apart from the ideal contribution to the equilibrium pressure, a term proportional to the midpoint number density, one also finds an electrostatic contribution that is due to the interaction between the monopolar and the quadrupolar part of the counterion charge density across the midplane and is proportional to the square of the midpoint density, see (\ref{quadPB1}). The attractive part thus does not look like bridging which has its origin in the connectivity and flexibility  of the polyelectrolyte chain, but more like a virial expansion in terms of the multipolar interactions. This is of course only true for point-like counterions. For extended counterions a bridging interpretation would be more appropriate as was already clear in the early studies of structured counterions \cite{forsman}.

\subsection{First order in $t$}

The above PB equation is a fourth order highly non-linear integro-differential equation and therefore difficult to handle
with conventional numerical procedures. Since here we are not particularly interested in the weak coupling results, we want only to show that the effect of quadrupolar moments, $t$, on the interaction pressure is to induce a small attractive contribution.
In order to show this we expand (\ref{quadPB1}) to the first order in $t$ and put $p=0$, which then reads
\begin{equation}
\quad \psi''=-C\rme^{-\psi}(1+\tfrac{1}{3}t\psi'^2-\tfrac{2}{3}t\psi'').
\label{PB1}
\end{equation}
This equation can be solved much more easily with conventional numerical methods. Although we must be aware that it is valid only for small $t$. The corresponding formula for interaction pressure in the 1st order in $t$ expansion reads
\begin{equation}
\tilde P=\tfrac12 C\,\rme^{-\psi(0)}.
\label{p1}
\end{equation}
Due to a convenient cancellation the formula contains no explicit $t$-dependence, but the pressure still depends on $t$ implicitly, due to the $t$ dependent potential $\psi$. As can be shown numerically for small $t$ the correction to the standard PB pressure depends linearly on $t$.
% MK: removed part B

\subsection{Numerical results in the mean-field limit}

Here we are not interested in the details of the mean-field results--they were analyzed in detail before \cite{bohinc1, bohinc,woon}-- but list them for completeness anyhow. In what follows we delimit ourselves  to counterions that posses monopolar and quadrupolar charge, so we do not take into account any dipoles ($p=0$).

\begin{figure*}[ht]
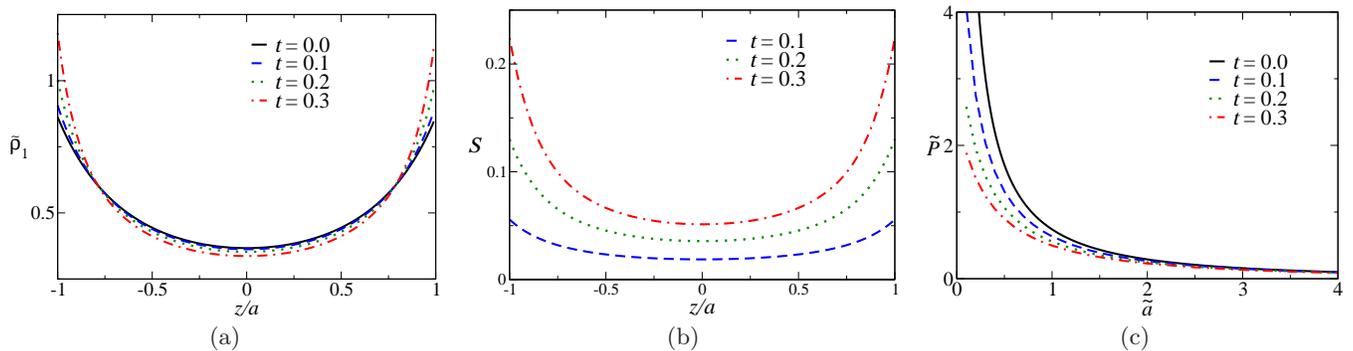
\begin{center}
\begin{minipage}[b]{0.32\textwidth}\begin{center}
\includegraphics[width=\textwidth]{WCrho.eps} (a)
\end{center}\end{minipage} \hskip0.25cm
\begin{minipage}[b]{0.32\textwidth}\begin{center}
\includegraphics[width=\textwidth]{S-WC.eps} (b)
\end{center}\end{minipage} \hskip0.25cm
\begin{minipage}[b]{0.31\textwidth}\begin{center}
\includegraphics[width=\textwidth]{WCpress.eps} (c)
\end{center}\end{minipage}
\caption{Mean-field  results for various $t$ in first-order approximation, (\ref{PB1}). (a) Counterion density profile, (\ref{quadPB1}). By increasing $t$ in the mean-field limit counterions are depleted from the center toward both surfaces. All the densities are rescaled to correspond to the same area under the curve. (b) Order parameter profile, evaluated from (\ref{eqx2}) and (\ref{eqS}). (c) Dimensionless pressure-distance curves obtained from the first-order approximation of (\ref{PB1}) and (\ref{p1}). Quadrupolar contributions are attractive.}
\label{figPB}
\end{center}
\end{figure*}
The density profile on Fig.~\ref{figPB}(a)  corresponds to monopolar density $\rho_1(z)$, (\ref{quadPB1}). Counterions are localized preferentially in the vicinity of both surfaces due to mutual repulsion that prevents them from being localized at the center. What we then observe is that with increase of the quadrupolar moment, $t$, counterions concentrate at surfaces even more. This can be easily explained. The potential energy of every quadrupolar particle in electrostatic potential is
\begin{equation}
\beta W=t\psi''\cos^2\theta.
\label{Wt}
\end{equation}
Since the second derivative of the mean-field potential $\psi''$ in the symmetric case considered here is a concave function of the coordinate $z$, this means that the quadrupolar force, $F=-\partial W/\partial z$, acts away from the center toward both surfaces.
%Corresponding MC simulations for $\Xi=0.1$ and $\Delta$ show a great importance of dielectric repulsion of counterions from both surfaces that change a density profile at their vicinities, whereas the rest of the profile in intermediate region is well predicted by WC theory. Note that densities are rescaled to correspond to the same area under the curve. As we shall see the behaviour at very small separations from the surfaces is mostly governed by SC approach.

The insight into the orientation of quadrupolar moments can be obtained through the average of the square of variable $x$, namely $\left<x^2\right>$,
\begin{equation}
\left<x^2\right>=\frac{\int x^2 u(x) \rmd x}{\int u(x) \rmd x}.
\label{eqx2}
\end{equation}
It is instructive to define an orientational order parameter
\begin{equation}
S={\textstyle\frac{1}{2}}\left(3 \left<x^2\right> -1\right).
\label{eqS}
\end{equation}
This order parameter can have values in the range from $-1/2$ to $1$. In a totally disordered system where quadrupoles point randomly in all possible directions $\left<x^2\right>=1/3$, and the orientational order parameter vanishes, $S=0$. In the completely ordered  case, $S=1$ and $\left<x^2\right>=1$, so that all quadrupolar moments are parallel with the $z$-axis and perpendicular to the walls, $\theta=0$. The other extremum, $S=-1/2$, corresponds to $\left<x^2\right>=0$ so that all quadrupolar moments are perpendicular to $z$-axis and parallel with the walls, $\theta=\pi/2$.

Fig.~\ref{figPB}(b) shows the numerically obtained profile of the orientational order parameter. It increases with increasing quadrupolar strength $t$ and is larger at both surfaces than at the center. The order parameter $S>0$ indicates that quadrupolar moments are preferentially aligned parallel to the $z$-axis. According to (\ref{Wt}) the electrostatic energy can be minimized by increasing $\cos^2\theta$, since $\psi''<0$. So $\left<\cos^2\theta\right>$ really is preferably increased above the value of $1/3$.

The quadrupolar contribution to the pressure for $t>0$ turns out to be to be attractive, Fig.~\ref{figPB}, as was shown before \cite{bohinc1, bohinc,woon}. Again this attractive contribution is obtained here on the mean-field level where the inter-counterionic correlations, essential in the strong coupling limit, play no role. It is indeed the {\em intra-counterion correlations} (entering here {\em via} the rigid structure assumed for counterions) that lead to such attractive contributions.  At this juncture it does not make much sense to us in continuing the multipolar expansion to yet higher orders, that would become relevant for even more aspherical and elongated charge distributions that would obviously entail also some molecular flexibility. In that case one could use the well developed theory of polyelectrolyte mediated interactions with much better confidence \cite{rudirev}.

\section{Strong Coupling Limit}

In the strong coupling limit, $\Xi\gg 1$, the system is highly charged and counterions are highly correlated.
The partition function (\ref{funint}) can be approximated by the two lowest order terms in the virial expansion with respect to the (renormalized) fugacity \cite{Netz}
\begin{equation}
Z = Z_{\rm SC}^{(0)} + \lambda' Z_{\rm SC}^{(1)} + {\cal O}(\lambda'^2).
\label{funint-SC}
\end{equation}
The second term corresponds to a one-particle partition function as the SC limit is effectively a single particle theory.
$Z_{\rm SC}^{(0)}$ and $Z_{\rm SC}^{(1)}$ are then given by
\begin{equation}
Z_{\rm SC}^{(0)} = \exp\left[- {\textstyle \frac12} \beta \!\!\int \rmd\Av r \rmd\Av r' \rho_0(\Av r) u(\Av r,\Av r') \rho_0(\Av r') \right],
\label{funint-SC-0}
\end{equation}
which is the exponential of the interaction between bare surface charges, and
\begin{eqnarray}
\frac{Z_{\rm SC}^{(1)}}{Z_{\rm SC}^{(0)}} &=&  \!\!\int\!\!\!\!\int \!\!\rmd\Av R \rmd\bomega \exp\biggl[- \beta\!\! \int\!\!\!\!\int\!\! \rmd\Av r \rmd\Av r' \hat \rho(\Av r; {\Av R}, \bomega) u(\Av r,\Av r') \rho_0(\Av r') -\nonumber\\
&&- {\textstyle\frac12} \beta \int\!\!\!\!\int \rmd\Av r \rmd\Av r' \hat \rho(\Av r; {\Av R}, \bomega) u(\Av r,\Av r') \hat \rho(\Av r'; {\Av R}, \bomega)\biggr],
\label{funint-SC-1}
\end{eqnarray}
which is the partition function of all possible (single) counterion configurations. The interaction potential in this part of the partition function is composed of the direct and image electrostatic interactions  (\ref{decomp-u}).

In the case of two charged surfaces with uniformly smeared surface charge density (\ref{sigmas}), the surface charge electrostatic potential does not depend on the $z$-coordinate; it is spatially homogeneous and is given by
\begin{equation}
\int \rmd\Av r' u(\Av r,\Av r')\rho_0(\Av r')=-\frac{\sigma a}{\varepsilon\varepsilon_0}.
\label{funint-u-00}
\end{equation}
The corresponding potential energy of a counterion in this surface electrostatic potential is given by
\begin{equation}
\int\!\!\!\!\int \rmd\Av r \rmd\Av r' \hat \rho(\Av r; {\Av R}, \bomega)u(\Av r,\Av r')\rho_0(\Av r')=-\frac{\sigma a}{\varepsilon\varepsilon_0}e_0q.
\label{funint-u-0}
\end{equation}

Since all the terms in density operator (\ref{chaden}), except the first one, depend on the gradients, {\em i.e.}, spatial derivatives, the counterion energy in a homogeneous external electrostatic potential depends only on the first, monopolar term (\ref{funint-u-0}). That means that higher multipoles do not interact directly with planar surface charge, but we must emphasize that this is not the case in inhomogeneous potential at curved surfaces, {\em e.g.}, at cylindrical or spherical surfaces or inhomogeneously charged surfaces.

As for the self-energy contribution, the second term in the exponent of (\ref{funint-SC-1}), it only picks up terms from the $z$-dependent parts of the image self-interaction, $u_{\rm im}$. The direct self-interaction, $u_0$, does not depend on coordinates so it can be discarded. The self-image energy is then
\begin{equation}
\int\!\!\!\!\int \rmd\Av r \rmd\Av r' \hat \rho(\Av r; {\Av R}, \bomega) u_{\rm im}(\Av r,\Av r') \hat \rho(\Av r'; {\Av R}, \bomega) =
\sum_{i=1}^9 w_i(z,\cos\theta).
\label{funint-u-1}
\end{equation}
Self-image contributions are among monopolar, dipolar and quadrupolar moments of the counterions interacting with their own electrostatic images of monopolar, dipolar and quadrupolar moments. Summing up all these contributions we remain with  nine terms of which only six are different. Writing them up {\em in extenso} we obtain
\begin{eqnarray}
w_1(z,\cos\theta)&=&q^2e_0^2\, u_\textrm{im}(\Av R,\Av R),\nonumber\\
w_2(z,\cos\theta)&=&qe_0p_0\, (\Av n\cdot\nabla)u_\textrm{im}(\Av r,\Av R)\big\vert_{\Av r=\Av R},\nonumber\\
w_3(z,\cos\theta)&=&qe_0t_0\, (\Av n\cdot\nabla)^2u_\textrm{im}(\Av r,\Av R)\big\vert_{\Av r=\Av R},\nonumber\\
w_4(z,\cos\theta)&=&w_2(z,\cos\theta),\nonumber\\
w_5(z,\cos\theta)&=&p_0^2\, (\Av n\cdot\nabla)(\Av n\cdot\nabla')u_\textrm{im}(\Av r,\Av r')\big\vert_{\Av r'=\Av r=\Av R},\nonumber\\
w_6(z,\cos\theta)&=&p_0t_0\, (\Av n\cdot\nabla)(\Av n\cdot\nabla')^2 u_\textrm{im}(\Av r,\Av r')\big\vert_{\Av r'=\Av r=\Av R},\nonumber\\
w_7(z,\cos\theta)&=&w_3(z,\cos\theta),\nonumber\\
w_8(z,\cos\theta)&=&w_6(z,\cos\theta),\nonumber\\
w_9(z,\cos\theta)&=&t_0^2\, (\Av n\cdot\nabla)^2(\Av n\cdot\nabla')^2u_\textrm{im}(\Av r,\Av r')\big\vert_{\Av r'=\Av r=\Av R}.\nonumber\\
\label{logatico}
\end{eqnarray}
Here, the first expression, $w_1$, corresponds to the interaction between a monopole and its own monopole image, therefore  it is proportional to $q^2$. The second term, $w_2$, corresponds to the monopole-dipole image interaction which is obviously the same as the dipole-monopole image interaction, $w_4$, and so on for all the higher order terms.

These image self-interaction terms can be expressed in a dimensionless form as
\begin{equation}
\tilde w_i(\tilde z,\cos\theta)={\textstyle\frac{1}{2}}\beta\, w_i(z,\cos\theta).
\end{equation}
The above expressions were derived for the planar geometry. Similar expressions but with different image potentials can be derived also in other geometries with dielectric inhomogeneities. Note that if there is no dielectric discontinuity, so that $u_\textrm{im}(\Av r,\Av r') = 0$, they are all identically zero, $w_i$=0!

Thus one can conclude at this point that in the dielectrically homogeneous case the higher order multipoles are completely irrelevant on the SC level. Only the monopolar term survives in the partition function. This is probably one of the most important conclusions of this work, so let us reiterate it: in a dielectrically homogeneous case in plan-parallel geometry the structure of the counterions as codified by their multipolar moments plays absolutely no role in the strong coupling limit!

For our case of two charged planar surfaces we evaluate the above expressions by using the Green's function (\ref{green}), which was derived for planar geometry with a step-function dielectric profile at the two boundaries $z = \pm a$. We get the following somewhat cumbersome expressions for the dimensionless self-image interactions defined above
\begin{widetext}
\begin{eqnarray}
\tilde w_1(\tilde z,x)&=&\frac{1}{2}\Xi\sum_{n\textrm{ odd}}\frac{n\tilde a\Delta^n}{(n\tilde a)^2-\tilde z^2}-\frac{\Xi}{4\tilde a}\,\ln(1-\Delta^2),\\
\tilde w_2(\tilde z,x)&=&\frac{1}{2}\Xi p\,x\tilde z\sum_{n\textrm{ odd}}\frac{n\tilde a\Delta^n}{[(n\tilde a)^2-\tilde z^2]^2},\nonumber\\
\tilde w_3(\tilde z,x)&=&-\frac{1}{8}\Xi t(1-3x^2)\left\{
\sum_{n\textrm{ even}}\frac{\Delta^n}{(n\tilde a)^3}+\sum_{n\textrm{ odd}}n\tilde a\frac{(n\tilde a)^2+3\tilde z^2}{[(n\tilde a)^2-\tilde z^2]^3}\,\Delta^n\right\},\nonumber\\
%\tilde w_4(\tilde z,x)&=&\tilde w_2(\tilde z,x),\nonumber\\
\tilde w_5(\tilde z,x)&=&\frac{1}{8}\Xi p^2\left\{(1-3x^2)\sum_{n\textrm{ even}}\frac{\Delta^n}{(n\tilde a)^3}+(1+x^2)\sum_{n\textrm{ odd}}n\tilde a
\frac{(n\tilde a)^2+3\tilde z^2}{[(n\tilde a)^2-\tilde z^2]^3}\,\Delta^n\right\},\nonumber\\
\tilde w_6(\tilde z,x)&=&\frac{3}{4}\Xi p t(1+x^2)x\,\tilde z\sum_{n\textrm{ odd}}
n\tilde a\frac{(n\tilde a)^2+\tilde z^2}{[(n\tilde a)^2-\tilde z^2]^4}\Delta^n,\nonumber\\
%\tilde w_7(\tilde z,x)&=&\tilde w_3(\tilde z,x),\nonumber\\
%\tilde w_8(\tilde z,x)&=&\tilde w_6(\tilde z,x),\nonumber\\
\tilde w_9(\tilde z,x)&=&\frac{3}{32}\Xi t^2\left\{(3-30x^2+35 x^4)\sum_{n\textrm{ even}}\frac{\Delta^n}{(n\tilde a)^5}+(3+2x^2+3x^4)\sum_{n\textrm{ odd}}n\tilde a
\frac{(n\tilde a)^4+10(n\tilde a\tilde z)^2+5\tilde z^4}{[(n\tilde a)^2-\tilde z^2]^5}\,\Delta^n\right\}.\nonumber
\end{eqnarray}
\end{widetext}
All indices $n$ here run only through positive values. Note again that self-image contributions, $\tilde w_i$, vanish when $\Delta=0$. Taking all this into account, the strong coupling interaction free energy, $\beta F=-k_{\mathrm{B}}T\,\ln\,Z_{\rm SC}^{(1)}$, {\em i.e.}, the part of the free energy that depends on the intersurface separation, can be written  in dimensionless form as
\begin{equation}
\tilde F/\tilde A=2\tilde a-2\,\ln\int_{-\tilde a}^{\tilde a}\rmd\tilde z\int_{-1}^1 \rmd x\,
\exp\Bigl(-\sum_{i=1}^9\tilde w_i(\tilde z,x)\Bigr).
\label{FreeEn}
\end{equation}

Just as in the WC case we have again assumed that the counterions are point-like particles so that the characteristic function $\Omega$ does not depend on their coordinate $\tilde z$ and therefore the integration goes from $-\tilde a$ to $\tilde a$. This means that we also disregard the possible entropic effects due to the finite size and anisotropy in the shape of counterions. These entropic contributions to the partition function are relevant only for intersurface separation on the order of the size of the counterion or smaller. In that regime of separations other, much stronger effects would come into play and compete with ionic finite size effects, thus these type of effects are not the focus of this paper.

The corresponding dimensionless pressure is simply obtained by taking the derivative of the free energy (\ref{FreeEn}) with respect to wall separation, $\tilde D = 2\tilde a$,
\begin{equation}
\tilde P=-\frac{\partial \tilde F/\tilde A}{2\,\partial \tilde a}.
\end{equation}

It is again obvious from here that in the dielectrically homogeneous case, {\em i.e.}, the case with no electrostatic images, where $w_{i}(\tilde z, x) = 0$, the strong coupling limit is given {\em exactly} by the monopolar term (the first term in (\ref{funint-SC-1})). Thus without the images we remain with the same form of the strong coupling interaction free energy as in the case of monopolar point charges. It is given by
\begin{equation}
\tilde F/\tilde A = 2 \tilde a - 2 \,\ln{\, \tilde a}\label{strongc2},
\end{equation}
with the corresponding pressure  $$ \tilde P = \frac{1}{\tilde a} - 1.$$

The higher order multipoles thus make {\em no} direct contribution to the strong coupling interaction free energy or forces between the bounding charged surfaces without dielectric inhomogeneities.

This is a very powerful result whose ramifications in fact impose rather stringent limits on the significance of the multipolar expansion of counterionic charge. It appears that for highly charged counterions, that are in fact the only ones where the multipolar expansion really makes sense, most of the electrostatics is properly captured by the monopolar term. Higher order multipoles simply do not contribute to the pressure in the system in the SC limit. Of course all of this is valid in the limit of homogeneous dielectric properties for planar surfaces without any image effects.

The counterion density profile can be extracted from (\ref{FreeEn}) as an integrand in the second term,
\begin{equation}
\tilde \rho(\tilde z)\propto\int_{-1}^1 \rmd x\,
\exp\Bigl(-\sum_{i=1}^9\tilde w_i(\tilde z,x)\Bigr).
\label{DensSC}
\end{equation}
In the case without the dielectric mismatch the density profile is simply homogeneous, $\rho(z)=\textrm{const}$.
Dielectric images induce an additional repulsion between the counterions and the surface charges pushing them towards the midplane region between the two charged bounding surfaces.

Orientational order of quadrupoles can be inspected {\em via} averaged $\left<x^2\right>$ which is here defined as
\begin{equation}
\left<x^2\right>=\frac{\int x^2 \exp\Bigl(-\sum_{i=1}^9\tilde w_i(\tilde z,x)\Bigr)\rmd x}{\int \exp\Bigl(-\sum_{i=1}^9\tilde w_i(\tilde z,x)\Bigr)\rmd x}.
\end{equation}
We can now use the same definition of order parameter $S$ as in WC, (\ref{eqS}).

% MK: removed part A (at the end of file)

\subsection{Numerical results in the SC limit}

We now present some numerical results for the  SC limit. Without any dielectric mismatches the density profile of the monopolar counterions is homogeneous, {\em i.e.}, it does not depend on $z$, which is a well known result \cite{Netz}. In the case of multipoles with dielectric images this is no longer true, but the counterion density does remain an even function of $z$.

\begin{figure*}[t]
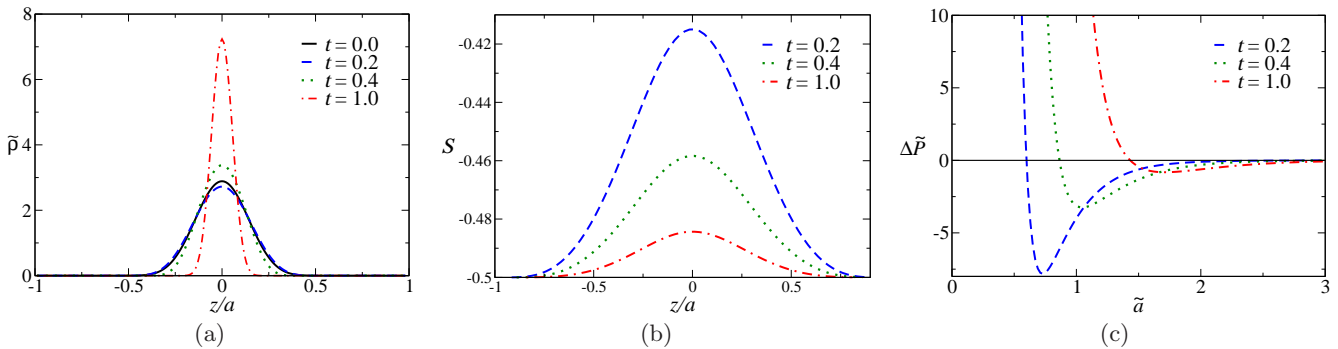
\begin{center}
\begin{minipage}[b]{0.302\textwidth}\begin{center}
\includegraphics[width=\textwidth]{rho50.eps} (a)
\end{center}\end{minipage} \hskip0.25cm
\begin{minipage}[b]{0.32\textwidth}\begin{center}
\includegraphics[width=\textwidth]{S-SC.eps} (b)
\end{center}\end{minipage} \hskip0.25cm
\begin{minipage}[b]{0.32\textwidth}\begin{center}
\includegraphics[width=\textwidth]{pressSC.eps} (c)
\end{center}\end{minipage}
\caption{Strong coupling results for $\Xi=50$ and $\Delta=0.95$ and various $t$. (a) Counterion density profile.
(b) Order parameter profile. Counterions are less ordered at the center. Order is increased with increasing $t$.
(c) Quadrupolar pressure contribution, $\Delta \tilde P=\tilde P_{t}-\tilde P_{t=0}$. Contribution is repulsive at small separations and becomes attractive at larger separations. At very large separations it goes to zero.}
\label{figSC}
\end{center}\end{figure*}
On Fig.~\ref{figSC} we show the numerical results for coupling parameter strength $\Xi=50$, dielectric jump $\Delta=0.95$ and $\tilde a=1$. Contrary to the WC situation, here counterions are localized in the vicinity of the midpoint between the two bounding surfaces due to image repulsion. Even without quadrupolar contributions ($t=0$) the monopolar interaction, $w_1$, is repulsive and long ranged, $w_1\sim z^{-1}$ \cite{matej1}, compressing the counterions to the midpoint.

When the quadrupolar parameter $t$, is increased a bit the midpoint peak first starts to widen.
This is due to the attractive monopole-quadrupole image  contribution, represented by the $w_3$ and $w_7$ terms which are proportional to $t$. As $t$ increases further (approximately for $t>8\tilde a^2/45$ for large $\Delta\Xi$) the repulsive quadrupole-quadrupole image contribution, $w_9$, which is proportional with $t^2$, takes over and confines counterions to the center even more so that the density peak  sharpens up. %All this is validated also by extensive MC simulations.

%We must emphasize here that this contribution in SC comes solely from images. Here we have three different contributions ($p=0$, $t>0$). The first one is $w_1$ term in eq. (\ref{FreeEn}) which is a repulsive monopole-image monopole intraction. This term has the longest range and has the greates contribution for reasonable quadrupolar moment ($t<1$). The second contribution comes from  dipole-image monopole and similarly monopole-image dipole that is captured in equal terms $w_3$ and $w_7$. This contribution is attractive since freely rotateble quadruple experience the attraction with point charge. Finally, the third contribution comes from quadrupolar-image quadrupolar moment, which is repulsive and is represented by term $w_9$.
%The importance of various terms depends on the wall separations. At large separations, $a$, the $w_1$ terms dominates and multipolar contributions can be neglected. At the same time the SC approximation can breaks down at too large separations.
%At very small separation the quadrupole-quadrupole, $w_9$, contribution has the greatest impact on density profile.

For the value $\Delta\Xi$ large enough, so that density profile becomes bell-shaped, the SC density profile can be approximated by a gaussian curve with a variance, {\em i.e.}, peak width, of the form
\begin{equation}
\delta \tilde z^2\simeq \frac{1}{\Delta\Xi}\frac{\tilde a^7}{\tilde a^4-3\tilde a^2 t+\frac{135}{16}t^2}.
\end{equation}
As can be seen, small $t$ widens the density peak, but further increasing $t$ over the value $t>8\tilde a^2/45$ the peak narrows. Also by increasing parameters $\Xi$, $\Delta$ the midpoint density peak narrows. This also means that the repulsive pressure between the walls increases. As the density peak narrows and the density of the counterions at the midpoint plane increases, the counterions become more ordered in the quasi two-dimensional midpoint layer, implying also that the correlations are stronger and therefore the SC limit should {\em a fortiori} be even more appropriate.

Fig.~\ref{figSC}(b) shows also the orientational order parameter for the SC limit. Contrary to the WC results, here the orientational order parameter is negative, implying that quadrupoles are preferably aligned perpendicular to $z$-axis. The parameter reaches its extremum $-1/2$ at both walls where quadrupoles are perfectly perpendicular to $z$-axis and parallel to the walls. This is due to the strong quadrupole-quadrupole image repulsion, $w_9$, which has its minimum at $x=0$ and is not a purely steric effect as in simulations of long charged stiff rods \cite{jan}.

Since monopole-monopole image  interaction, $w_1$, has the longest range, we can expect that for large separations between the walls, $2\tilde a$, this interaction will dominate and higher multipoles can be again neglected. Fig.~\ref{figSC}(c) shows the pressure difference  $\Delta \tilde P$, which is the difference between the pressure with finite $t$ and the pressure without the quadrupolar contributions, $t=0$. As expected this difference indeed goes to zero at large separations. At moderate separations the quadrupolar contribution becomes negative, which is caused by monopole-quadrupole image attractions, $w_3$ and $w_7$, that are shorter ranged than $w_1$. At even smaller separations quadrupole-quadrupole image repulsion, $w_9$, becomes dominant, since it has the shortest range and varies approximately as $\sim a^{-5}$. The larger the $t$ the smaller is thus the maximal quadrupolar attraction at larger distance since it is overwhelmed by repulsive contributions.

\section{Multipoles, correlations, bridging and all that jazz}

There appear some similarities between the polyelectrolyte bridging interaction \cite{licer,podgornik-rev}  and the attraction seen with strongly charged counterions. Recent simulations and density functional results \cite{jan} in fact do accentuate a close connection between polyelectrolyte bridging and strong-coupling electrostatics, a line of reasoning that we will explore here in finer detail.

Turesson {\em et al.} \cite{jan} indeed find out that for flexible polyelectrolytes bridging attraction appears to be the dominant source of attractive interaction. On the other hand, for stiff charged rods the strong adsorption to the bounding charged walls prevents  bridges to form. Instead a very strong correlation attraction is apparent at shorter separations. Flexibility thus extends the range of attractive interactions as their nature changes from short range correlation attraction to longer ranged polyelectrolyte bridging.  Similar conclusions have been also reached in \cite{jason}.

Here we will formulate the exact correspondence between the intuitive notion of ``bridging'' in the case of multivalent counterions and the definition of ``bridging interaction'' in the case of charged polymers. Let us analyse first the partition function for two disparate systems: a counterion Coulomb fluid confined between two strongly charged walls, and a flexible polyelectrolyte chain confined in the same geometry with weakly charged walls. For point counterions between two charged surfaces the grand canonical  partition function in the SC limit \cite{netz-a} can be written to the first order in fugacity  as
\begin{equation}
Z = Z_{\rm SC}^{(0)} + \lambda'~ Z_{\rm SC}^{(1)} + \dots,
\end{equation}
where
\begin{eqnarray}
Z_{\rm SC}^{(0)} = {\mathcal C} \int {\cal D}[\phi(\Av r)]~ \exp\biggl(&-&{\ul12} \beta\varepsilon_0 \int \varepsilon({\bf r})(\bnabla \phi({\bf r}))^2 \rmd{\bf r} + \nonumber\\
&+& \rmi \beta \int \rho_0({\bf r}) \phi({\bf r}) \rmd{\bf r} \biggr),
\end{eqnarray}
and
\begin{eqnarray}
Z_{\rm SC}^{(1)}&=& \!{\mathcal C} \!\!\int \!\!\rmd{\bf R}\,\Omega(\Av R)\!\!\!\int\!\! {\cal D}[\phi(\Av r)] \exp \biggl(\!\!-{\ul12} \beta \varepsilon_0 \!\!\int \!\!\varepsilon({\bf r})(\bnabla \phi({\bf r}))^2 \rmd{\bf r} +\nonumber\\
&+&\rmi \beta \!\int \!\rho_0({\bf r}) \phi({\bf r}) \rmd{\bf r} + \rmi \beta qe_0 \! \int\! \rho(\Av r - \Av R) \phi({\bf r}) \rmd{\bf r}\biggr).
\label{comp-1}
\end{eqnarray}
The above partition function is of course exactly the same as the one in (\ref{funint-SC}) after one evaluates the functional integrals indicated above explicitly.  Here $\rho_0({\bf r})$ is again the external fixed charge on the two bounding surfaces located at
$ z = \pm \ul a2$ and the shorthand $\rho(\Av r - \Av R)$ is introduced for $\rho(\Av r - \Av R) = \delta^3(\Av r - \Av R)$. 

On the other hand the partition function for a flexible polyelectrolyte chain is given by \cite{Podgornik1}
\begin{eqnarray}
Z_{\mathrm{P}} = &{\mathcal C}&\!\! \int\!\! {\cal D}[\phi(\Av r)] ~\exp\biggl(\!\!-{\ul12} \beta \varepsilon_0\!\! \int \!\! \varepsilon({\bf r})(\bnabla \phi({\bf r}))^2 \rmd{\bf r} +\\
&+& \rmi \beta\! \int\!\! \rho_0({\bf r}) \phi({\bf r}) \rmd{\bf r} +  {\rm ln}{\int\!\!\!\!\int {\cal G}_{\phi}(\Av r, \Av r'; N) \,\rmd{\bf r}\rmd{\bf r'}}\biggr).\nonumber
\end{eqnarray}
where ${\cal G}_{\phi}(\Av r, \Av r')$ is the Green function of a polyelectrolyte chain in external electrostatic field $\phi$ given by \cite{Edwards}
\begin{eqnarray}
{\cal G}_{\phi}(\Av r, \Av r'; N) &=& \int_{\Av r= \Av R(0)}^{\Av r' = \Av R(N)} {\cal D}[\Av R(n)]\times\\
&&\hspace{-12ex}\times\exp\biggl(\!- {\textstyle{\frac{3}{2 \ell^2}}} \!\int_0^N \!\! \left( \frac{\rmd\Av R(n)}{\rmd n}\right)^2 \rmd n + \rmi\beta qe_0\! \int_0^N \!\!\phi(\Av R(n)) \rmd n\biggr),\nonumber
\end{eqnarray}
where $\Av R(n)$ is the coordinate of a polyelectrolyte segment $n$ and the boundary condition is set as
$\lim_{N \rightarrow 0} {\cal G}_{\phi}(\Av r, \Av r'; N) = \delta^3(\Av r - \Av r')$. Representing the Green function {\em via} a corresponding Edwards equation one can derive that the polyelectrolyte monomer density $ \rho_{\phi} (\Av r; N)$ is given by a functional derivative of ${\rm ln}{\int\!\!\!\!\int {\cal G}_{\phi}(\Av r, \Av r'; N) \,\rmd{\bf r} \rmd{\bf r'}}$ with respect to the fluctuating electrostatic potential $\phi(\Av r)$  \cite{Edwards}. The polyelectrolyte monomer density  has to satisfy the condition $\lim_{N \rightarrow 0} \rho_{\phi=0} (\Av r; N) = \delta^3(\Av r).$

In the planar geometry with two bounding surfaces, one can center the monomer density at the origin of the coordinate system, whose $z$ axis is directed along the surface normal and has the origin at the midplane. Denoting $\Av R_0 = (0, 0, 0)$ one can thus write $\rho_{\phi=0} (\Av r; N) = \rho_{\phi=0} (\Av r - \Av R_0; N)$. In the limit of small electrostatic potentials, the lowest order ({\em i.e.}, the
first-order) solution of the polyelectrolyte Green function can be written as \cite{Edwards}
\begin{equation}
{\rm ln}{\int\!\!\!\!\int{\cal G}_{\phi}(\Av r, \Av r'; N) \,\rmd{\bf r} \rmd{\bf r'}} \cong
\rmi \beta qe_0 \int \rho_{\phi=0} (\Av r - \Av R_0; N) \phi(\Av r) \rmd\Av r.
\label{green-2}
\end{equation}
By construction this solution clearly corresponds to a weak coupling of the polyelectrolyte to the external field generated by the fixed charges at the boundary of the system. For large values of this charge higher orders in the expansion (\ref{green-2}) would certainly have to be taken into account.

In the case of weak coupling the partition function of a polyelectrolyte chain is then given approximately by
\begin{eqnarray}
Z_{\mathrm{P}} &\cong& Z_{\rm WC}^{(0)}[\rho_{\phi=0} (\Av r - \Av R_0; N)] = \label{comp-2}\\
&=& {\mathcal C} \int {\cal D}[\phi(\Av r)] ~\exp\biggl(-{\ul12} \beta\varepsilon_0 \int \varepsilon({\bf r})(\bnabla \phi({\bf r}))^2 \rmd{\bf r} +\nonumber\\
&+& \rmi \beta \!\int \!\rho_{0}({\bf r}) \phi({\bf r}) \rmd{\bf r} +  \rmi \beta qe_0 \!\int \!\rho_{\phi=0} (\Av r - \Av R_0; N) \phi(\Av r) \rmd\Av r \biggr).\nonumber
\end{eqnarray}
If we now compare the expressions (\ref{comp-1}) and (\ref{comp-2}) we can establish the following identity
\begin{equation}
Z_{\rm SC}^{(1)} =  \lim_{N \rightarrow 0}\int \rmd{\bf R}\,\Omega(\Av R)~ Z_{\rm WC}^{(0)}[\rho_{\phi=0} (\Av r - \Av R; N)],
\end{equation}
with the simple (first order SC) counterion partition function on the l.h.s. and the polyelectrolyte (zeroth order WC) partition function on the r.h.s. This ``duality'' relation connects the partition function of a strongly coupled point counterion system with the partition function of a weakly coupled flexible polyelectrolyte system. Since $Z_{\rm WC}^{(0)}$ for finite $N$ describes polyelectrolyte bridging interactions it is clear that the above formula establishes a formal connection between strong coupling point counterion attractive interactions and weak coupling polyelectrolyte bridging interactions.

\section{Conclusions}
Using field-theoretic methods we derived a description of the counterion-mediated electrostatic interaction when the counterions posses internal degrees of freedom such as a rotational axis. We concentrate on the symmetrically charged planar surfaces with a dielectric mismatch on both sides and counterions with quadrupolar moments in between. We analyse this system in two different regimes, namely in the weak coupling (mean-field) and the strong coupling limit. The former one describes the case of low surface charge and low counterion valency, whereas the latter one describes high surface charge and high counterion valency.

In the mean-field limit, we derived the Poisson-Boltzmann equation for counterion density in the case that the counterions posses dipolar as well as quadrupolar moments. Since the equation is highly non-linear, we solved it numerically only in the first order approximation, which should be valid for small quadrupolar moments, $t$.
As is already known the dielectric mismatches play no role in the mean-field theory and have therefore no effect on quadrupoles \cite{matej1}. One could add those corrections in by hand \cite{onuki}, but such an approach does not strictly corresponds to the mean-field analysis. Quadrupolar interaction affects the counterion density distribution as well as interactions between the charged surfaces. Counterions are depleted from the central region and concentrated in the vicinity of both surfaces. The orientation of quadrupoles is preferentially  parallel to the $z$-axis as the moment $t$ is increased. The highest alignment is reached right next to both bounding surfaces. The quadrupolar contribution to the interaction pressure is attractive.

In the strong coupling limit higher order multipoles also play an important role in the intersurface interactions  but only for a dielectrically inhomogeneous case. Without any dielectric discontinuities in the system the higher multipoles simply do not matter in plan-parallel case and the monopolar part of the partition function captures all the strong coupling effects. This is a completely general result for plan-parallel surfaces and should be of some importance in assessing the electrostatically mediated forces between strongly charged planar surfaces. As already mentioned the multipoles can have effects near curved surfaces even without dielectric discontinuities.
If compared to the case of monopolar counterions, higher multipoles interact only {\em via} the orientationally dependent part of the image self-interaction. In our analysis we focused on the counterions with monopolar and quadrupolar moments, so that the interaction is composed of three contributions. These are the monopole-monopole image  interaction, which is long ranged and repulsive, the monopole-quadrupole image as well as quadrupole-monopole image  interactions that are shorter range and attractive. Finally, very short ranged quadrupole-quadrupole image contribution is repulsive and plays a dominant role at very small distances.

We are at present involved in extensive Monte Carlo simulations in order to explore the validity of the formal developments
described above. Preliminary results completely vindicate the theoretical analysis.

\section{Acknowledgement}

R.P. would like to acknowledge the financial support by the Slovenian Research Agency under contract
Nr. P1-0055 (Biophysics of Polymers, Membranes, Gels, Colloids and Cells). M.K. would like to acknowledge the financial support by the Slovenian Research Agency under the young researcher grant. A.N. would like
to  acknowledge financial support by the Institute for Research in Fundamental Sciences (IPM), Tehran. 
This research was supported in part by the National Science Foundation under Grant No. PHY05-51164 (while at the KITP program {\em The theory and practice of fluctuation induced interactions}, UCSB, 2008). Y.S.J. is grateful to M.W. Kim for useful discussions. Y.S.J. and P.A.P. were supported in part by the National Science Foundation (Grants DMR-0503347, DMR-0710521) and MRSEC NSF DMR-0520415. Y.S.J. and P.A.P. were supported by a Korea Science and Engineering Foundation (KOSEF) grant funded by the Korean Government (MEST)
(grant code: R33-2008-000-10163-0) and the Brain Korea 21 projects by the Korean Government.

\end{document}